\documentclass[sigconf]{acmart}
\usepackage{algorithm}
\usepackage{algorithmic}
\usepackage{multirow}
\usepackage{enumitem}
\setlist[itemize]{leftmargin=*}
\setlist[enumerate]{leftmargin=*}

\def\BibTeX{{\rm B\kern-.05em{\sc i\kern-.025em b}\kern-.08emT\kern-.1667em\lower.7ex\hbox{E}\kern-.125emX}}

\copyrightyear{2019}
\acmYear{2019}
\setcopyright{acmcopyright}
\acmConference[MM '19]{Proceedings of the 27th ACM International Conference on Multimedia} {October 21--25, 2019} {Nice, France}
\acmBooktitle{Processdings of the 27th ACM Int'l Conf. on Multimedia (MM`19), Oct. 21--25, 2019, Nice, France}
\acmPrice{15.00}
\acmDOI{10.1145/3343031.3350858}
\acmISBN{978-1-4503-6889-6/19/10}

\begin{document}
\fancyhead{}

\title{Personalized Hashtag Recommendation for Micro-videos}

\author{Yinwei Wei}
\affiliation{
    \institution{Shandong University}
}
\email{weiyinwei@hotmail.com}

\author{Zhiyong Cheng$^*$}
\affiliation{
    \institution{Shandong Computer Center (National Supercomputer Center in Jinan), Qilu University of Technology (Shandong Academic of Sciences)}
}
\email{jason.zy.cheng@gmail.com}

\author{Xuzheng Yu}
\affiliation{
    \institution{Shandong University}
}
\email{XuzhengYuuu@gmail.com}

\author{Zhou Zhao}
\affiliation{
    \institution{Zhejiang University}
}
\email{zhaozhou@zju.edu.cn}

\author{Lei Zhu}
\affiliation{
    \institution{Shandong Normal University}
}
\email{leizhu0608@gmail.com}

\author{Liqiang Nie$^*$}
\affiliation{
    \institution{Shandong University}
}
\email{nieliqiang@gmail.com}
\thanks{$^*$Corresponding author: Zhiyong Cheng and Liqiang Nie}
\begin{abstract}
    Personalized hashtag recommendation methods aim to suggest users hashtags to annotate, categorize, and describe their posts. The hashtags, that a user provides to a post (e.g., a micro-video), are the ones which \emph{in her mind} can well describe \emph{the post content where she is interested in}. It means that we should consider both \emph{users' preferences on the post contents} and \emph{their personal understanding on the hashtags}. Most existing methods rely on modeling either the interactions between hashtags and posts or the interactions between users and hashtags for hashtag recommendation. These methods have not well explored the complicated interactions among users, hashtags, and micro-videos.
    In this paper, towards the personalized micro-video hashtag recommendation, we  propose a Graph Convolution Network based Personalized Hashtag Recommendation (GCN-PHR) model, which leverages recently advanced GCN techniques to model the complicate interactions among <users, hashtags, micro-videos>  and learn their representations. In our model, the users, hashtags, and micro-videos are three types of nodes in a graph and they are linked based on their direct associations. In particular, the message-passing strategy is used to learn the representation of a node (e.g., user) by aggregating the message passed from the directly linked other types of nodes (e.g., hashtag and micro-video). Because  \emph{a user is often only interested in certain parts of a micro-video} and \emph{a hashtag is typically used to describe the part (of a micro-video) that the user is interested in}, we leverage the attention mechanism to filter the message passed from micro-videos to users and hashtags, which can significantly improve the representation capability.
    Extensive experiments have been conducted on two real-world micro-video datasets and demonstrate that our model outperforms the state-of-the-art approaches by a large margin.
\end{abstract}


%

\maketitle
\section{Introduction}
In social networks, such as Twitter\footnote{https://twitter.com/.} and Instagram\footnote{https://www.instagram.com/.}, hashtags are frequently used to annotate, categorize, and describe the posts according to users' preferences.
\begin{figure}
	\centering
	  \includegraphics[width=0.4\textwidth]{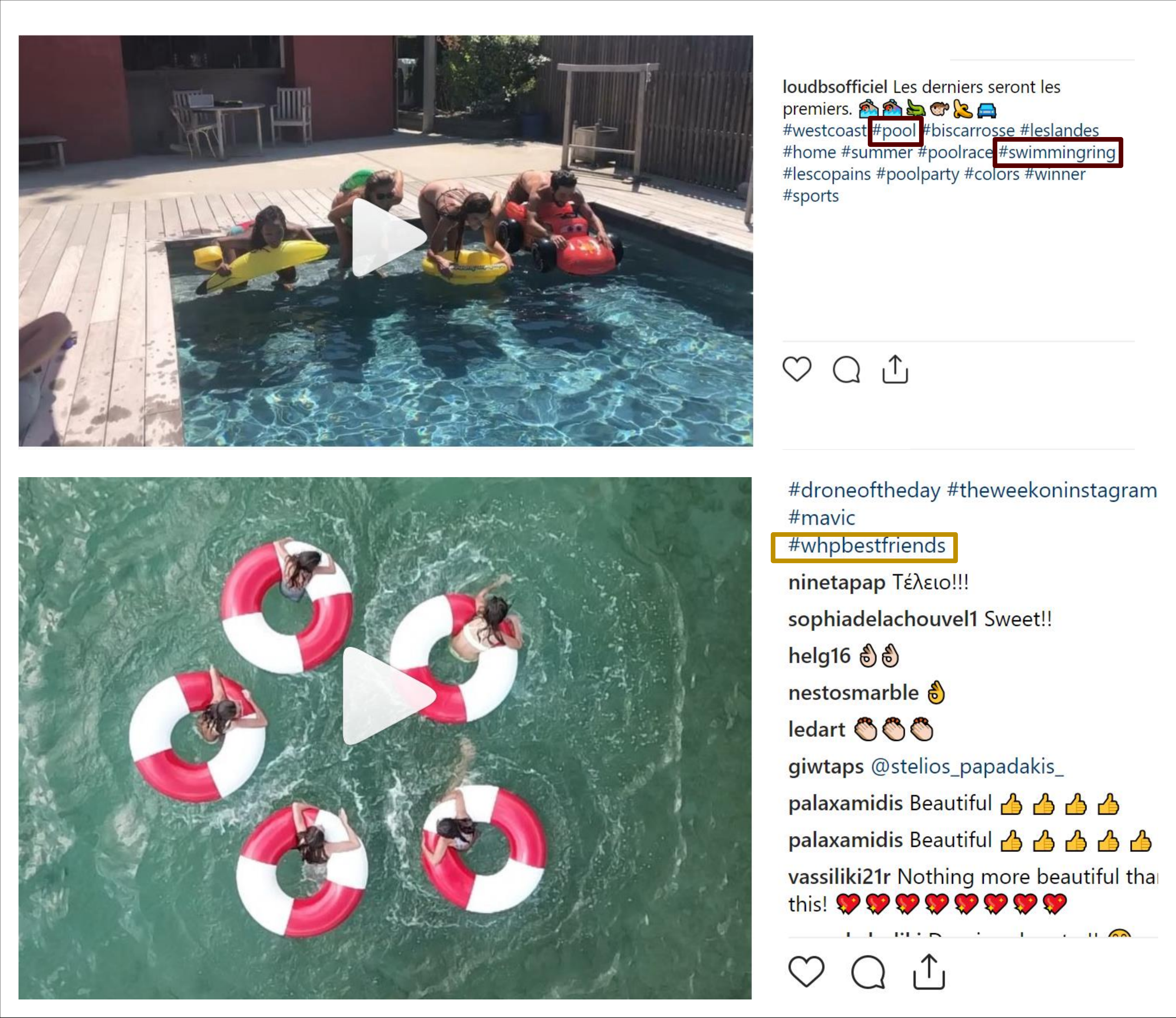}
	  \vspace{-1mm}
	  \caption{Exemplar demonstration of user preference. Two users publish similar micro-videos with different hashtags.}
	\label{fig_1}	
	\vspace{-6mm}
\end{figure}
They may consist of any arbitrary combination of characters led by a hash symbol `\#'~(e.g. \#Puppy and \#thegoodlife).
Hashtags are created by users and they hence can be treated as the self-expression of users, conveying the users' preferences on posts and their usage styles of hashtags.
With these hashtags, users can easily search and manage their historical posts and track others' posts.
Moreover, studies have shown that hashtags can provide valuable information about several tasks, such as sentiment analysis~\cite{wang2011topic} and video understanding~\cite{wang2012event}.
However, due to the inconvenient typing on smartphones, only sparse users are willing to provide hashtags to their posts~(e.g. micro-videos).
According to the statistics\footnote{https://www.omnicoreagency.com/instagram-statistics/.}, by September 2018, more than $33$ million micro-videos without any hashtag are uploaded per day on Instagram. Therefore, personalized hashtag recommendation has attracted considerable attention from industrial and academic communities.\par

In recent years, several methods have been proposed to automatically suggest appropriate hashtags to users rely on their posts' content. 
\begin{figure*}
	\centering
	  \includegraphics[width=0.90\textwidth]{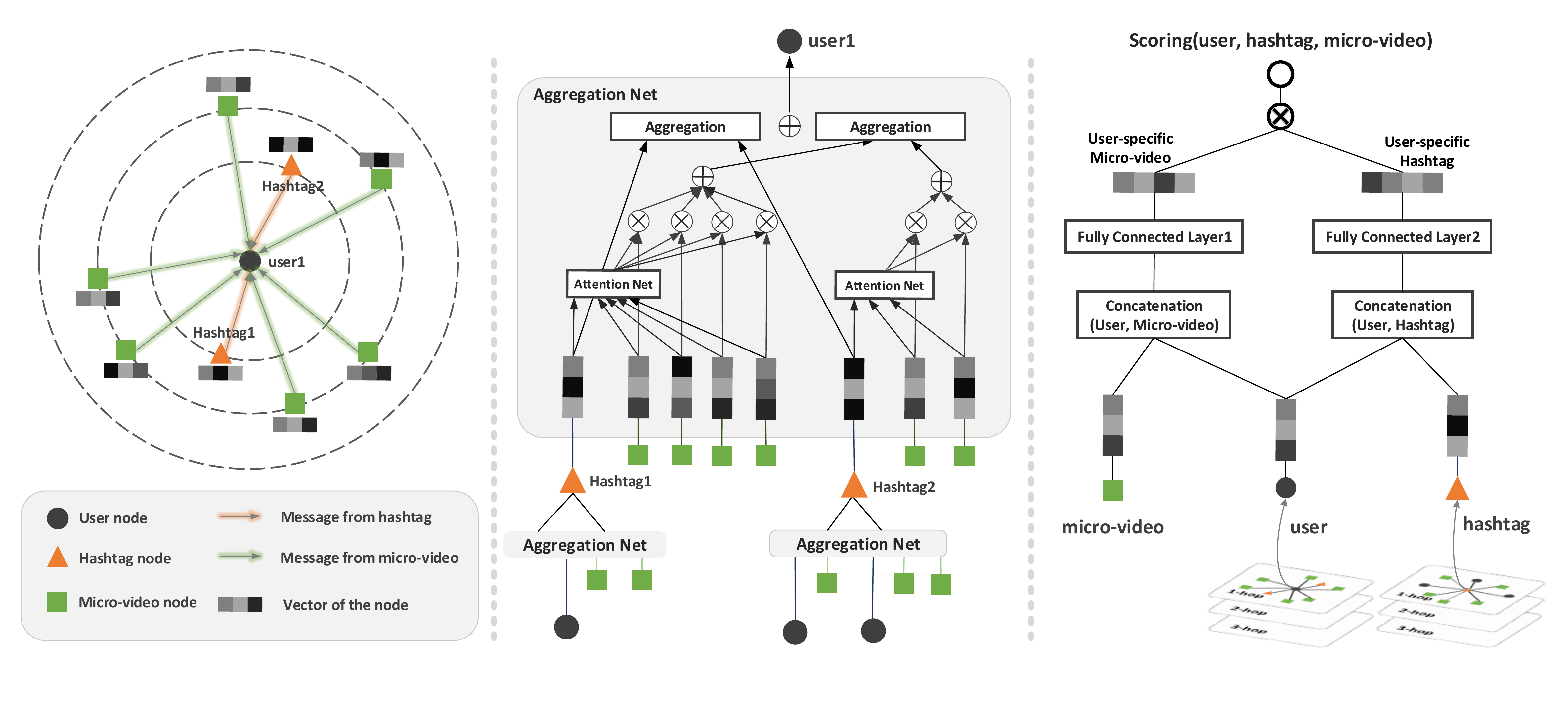}
	  \vspace{-9mm}
	  \caption{Schematic illustration of our proposed graph-based convolutional network. In this graph, we treat users, hashtags and micro-videos as nodes. We use two edges to link them if a user annotates a micro-video use a hashtag.  }
	\label{fig_2}	
	\vspace{-3mm}
\end{figure*}
Pioneer efforts~\cite{rawat2016contagnet,chunseong2017attend,denton2015user,tran2018hashtag}
view personalized hashtag recommendation as a multi-class classification or information retrieval problem to predict the hashtag.
For instance, Tran \textit{et al.}~\cite{tran2018hashtag} presented a hashtag recommendation method which leverages the historical tweets, used hashtags, and social interaction to profile the users, and then searched the similar users' tweets for recommending the hashtags.
Denton \textit{et al.}~\cite{denton2015user} used metadata~(e.g., age, gender, etc.) to characterize users, and integrated users' representations with features of their posts to infer the categories.
Nevertheless, these methods ignore the interactions between users and hashtags.
For example, the hashtag `\#rock' may be annotated to totally different images by music fans and mountaineers.
Users indeed always have their own preferences on hashtag usages.
Being aware of this problem, Alvari~\cite{alvari2017twitter} applied the matrix factorization based collaborative filtering~(CF) method to model the interactions between users and hashtags for personalized hashtag recommendation. Furthermore, considering the inherent subjective of the hashtag, Veit \textit{et al.}~\cite{veit2018separating} trained a user-specific hashtag model on image-user-hashtag triplets, by taking the hashtag usage patterns of users into account.  However, this method has not well exploited the post content, which contains rich information about user preference and hashtag semantics.


Notice that the CF-based methods can only linearly model interactions among users, hashtags, and posts. Therefore, they merely capture the user-specific hashtag usage patterns rather than the representation of the user preference and hashtag semantic, while the latter is the core of personalized hashtag recommendation.
The hashtag semantic should be consistent with the corresponding part of post content, which in turn is the interested part of the user. For example, as shown in Figure~\ref{fig_1}, different users may annotate different hashtags for similar micro-videos according to their preferences.
Considering this fact, the user preference and hashtag semantic representation actually have an complicated interaction with posts' features.
Therefore, how to exploit such complicated interactions to represent the user preference and hashtag semantic is nontrivial. The abundant and multimodal information of micro-videos makes hashtag personalized recommendation even more challenging~\cite{nie2016learning,Liu2017Towards,Nie2017Enhancing,8488491}. To deal with the aforementioned challenge, we propose a novel graph convolutional network~(GCN) based method to model such complicated interactions for micro-video hashtag personalized recommendation, as illustrated in Figure~\ref{fig_2}. In our model, the users, hashtags, and micro-videos are three types of nodes in a graph and they are linked based on their direct associations. In other words, user nodes are connected to their historical micro-video nodes and used hashtag nodes, and hashtag nodes are connected to their accompanied micro-video nodes and corresponding user nodes.
Based on this graph, our model learns the user and hashtag representations using the graph convolutional techniques.

According to the message-passing method~\cite{berg2017graph}, the representation of a node can be learned based on the message passed from its neighbor nodes. Based on this idea, in our model, the representation of a user node is learned by aggregating the message from her neighbor hashtag and micro-video nodes; and  the representation of a hashtag is learned based on the message from its neighbor user and micro-video nodes. As a user is often only interested in certain parts of a micro-video and a hashtag is typically used to describe a particular part of a micro-video, the message passed from a micro-video to its neighbor user or hashtag may contain redundant or noisy information.
To solve this problem, we employ the attention mechanism to filter the information related to the user and hashtag.
Specifically, our model uses the hashtag representation to filter the micro-video information in the corresponding user node representation, since hashtags are used by users to express their interests in the micro-video.
Analogously, the representation of user preferences are used to filter the micro-video information to model the hashtag semantics, because the user preferences can be used to identify which parts in the micro-video are tagged by the hashtags.
Based on this design, our model can achieve better user and hashtag representation learning. Thereafter, our model further learns the user-specific micro-video features and user-specific hashtag semantics with obtained representations of user preference and hashtag semantic, which are then used for personalized hashtag recommendation for micro-videos.
To verify the effectiveness of our model, we perform extensive experiments on two micro-video benchmark datasets.
The experimental results show that our proposed model outperforms several state-of-the-art approaches.\par
The main contributions of this work are threefold:
\vspace{-1mm}
\begin{itemize}
\item To the best of our knowledge, this is the first work which attempts to design a personalized hashtag recommendation method for micro-videos. Our model can comprehensively model the interactions between users, hashtags, and micro-videos for hashtag recommendation.

\item We design a novel GCN based hashtag recommendation method. In particular, we introduce the attention mechanism to filter the redundant message passed from micro-videos to users and hashtags in the graph, which can significantly enhance the learning of user and hashtag representations.

\item We conduct extensive experiments on two real-world micro-video datasets demonstrate the superiority of our method over several state-of-the-art methods. In addition, we released our codes, datasets, and parameters to facilitate related researches\footnote{https://github.com/weiyinwei/GCN\_PHR.}.
\end{itemize}

\section{Related Work}
In this section, we mainly review the studies that are most related to our work, including hashtag recommendation and graph convolutional network. \par

\subsection{Hashtag Recommendation}
Due to the widespread use of hashtags on various social platforms, hashtag recommendation has attracted increasing attention in recent years and many methods~\cite{CLEAR,XLCMM_IJCAI2017} have been proposed.
Those methods can be broadly grouped into two categories: 1) the ones based on the interactions between hashtags and posts; and 2) the ones based on the interactions between users and posts. \par

Methods in the first category attempt to learn the representation of hashtags based on their corresponding posts, and then recommend hashtags according to the post content. For example,  Zhang \textit{et al.}~\cite{zhang2014time} assumed that hashtag and textual content are two different descriptions of the same thing. Based on this assumption, they proposed a topical translation model, which extracts topics from contents and translates them to the hashtags, to recommend hashtags for microblogs. Dey \textit{et al.}~\cite{dey2017emtagger} regarded the hashtag recommendation as a word prediction task. Specifically, a hashtag is treated as a word in tweets and then the word-embedding techniques are applied to learning the representation of hashtags and predicting hashtags in tweets. Nevertheless, the aforementioned methods ignore the user-related factors, such as user preferences and language habits. In fact, some users may use different hashtags to describe the same content~(i.e., synonyms) and some may use the same hashtag to describe different contents~(i.e., polysemy). To take user preference into account, Veit~\textit{et al.}~\cite{veit2018separating} applied the three-way tensor product to learn user-specific representations for hashtags. The learned representations are then used to measure the similarity scores between hashtags and posts. However, this method merely captures user preferences on hashtags while ignores user interest in the post content. Different from these methods, our model learns both the user-specific hashtag representation and user-specific post representations, which are then used to compute the user-specific similarity score of each post and hashtag pair. \par
Methods in the second group treat the hashtag recommendation as a classification or prediction problem by learning the interactions between the user preference and post content. Denton \textit{et al.}~\cite{denton2015user} took the user demographic information into consideration (e.g., age, gender, etc.) to model user preferences, which are combined with image features and then mapped into the same embedding space with hashtags. In this space, the hashtags are matched with the user-image combination representations for the recommendation.  Rawat \textit{et al.}~\cite{rawat2016contagnet} fed the integration of user representation and image features into a deep neural network to predict hashtags for images. In this model, they considered the  contextual preferences of users on images~(e.g. time and geo-location) in user representation learning. Recently, Park \textit{et al.}~\cite{chunseong2017attend} modeled user preferences based on  their most frequently used hashtags in historical posts and devised a Context Sequence Memory Network for hashtag prediction based on user preferences and image features.  All the above methods model user preferences on the global level of a post, however, it is common that a user is interested in a particular part of a post and  provides a hashtag according to the interested part, especially for the posts with rich content (such as micro-videos).  In this work, we will model the user preference based on the content interests them in micro-videos and learn user-specific micro-video representations according to content that interests them. \par

\subsection{Graph Convolutional Networks}
Graph convolutional networks have been widely used in various applications, such as computer vision~\cite{li2018factorizable,Jain2015Structural,Garcia2018Few,Marino,Zhang}, disease or drug prediction~\cite{Ktena2017Distance,Zitnik2018Modeling,Parisot2017Spectral}, and Chemistry~\cite{li2018learning,fout2017protein,you2018graph,Gilmer2017Neural}. Due to their powerful capability in representation learning, GCNs have also been exploited to model the interactions between users and items~\cite{kipf2016semi,niepert2016learning,monti2017geometric,KGAT, NGCF} for recommendation. For example, a general inductive GCN framework GraphsSAGE~\cite{hamilton2017inductive}, which learns node representation based on both the topological structure of graph and node feature information, has been verified in citation recommendation and video recommendation.  However, GCN-based recommendation models often need to store the whole graph in GPU memory, which limits their applications in large-scale recommendation tasks with billions of items and millions of users. To tackle this issue, Ying \textit{et al.}~\cite{ying2018graph} proposed an efficient GCN algorithm PinSage which combines efficient random walks and graph convolutions to learn node representations. This method has been demonstrated to be scalable for large-scale recommendation tasks. From a different perspective, Berg \textit{et al.}~\cite{berg2017graph} considered recommendation as link prediction on graphs and proposed a novel graph auto-encoder framework named Graph Convolutional Matrix Completion~(GCMC). Based on a proposed message-passing model, it uses a graph convolution layer in the encoder to learn the representations of users and items and then leverages the representations to reconstruct the rating links through a bilinear decoder. \par

Different from the previous works which typically consider two types of nodes, we need to model the interactions among three kinds of nodes (i.e., user, hashtag, micro-video) in this work. As in GCMC, we also adopt the message-passing model to learn the interaction among the different types of nodes. The difference is that we propose to use the attention mechanism to filter the message from micro-videos to users and hashtags. To the best of our knowledge, this is the first work to use GCN-model for personalized hashtag recommendation and also the first attempt to apply GCN techniques for graphs with three types of nodes.

\section{Our Proposed Method}
\subsection{Problem Setting and Model Overview}
\subsubsection{Problem setting.} \label{sec:ps}
Before describing our method, we would like to introduce the problem setting first. Given a dataset with a micro-video set $\mathcal{V}$, a hashtag set $\mathcal{H}$, and a user set $\mathcal{U}$, in which a micro-video $v_k\in\mathcal{V}$ is uploaded by a user $u_i\in\mathcal{U}$ and some hashtags $h_j\in\mathcal{H}$ are provided by $u_i$ to $v_k$. Based on this dataset, the goal is to learn a personalized hashtag recommendation model, which could automatically recommend hashtags from $\mathcal{H}$ to a new micro-video $v$ uploaded by a user $u$. Because we would like that the recommended hashtags will be adopted by the user, the recommended hashtags should not only match the micro-video contents but also fit the personal preferences of users. To achieve this goal, we apply the graph convolutional networks to modeling the complex interactions among three types of entities: users, hashtags, and micro-videos.

Let $\mathcal{G}=(\mathcal{W}, \mathcal{E})$ be an undirected graph, where $\mathcal{W}$ denotes the set of nodes and $\mathcal{E}$ is the set of edges. Specifically, $\mathcal{W}$ consists of three types of nodes, which represents the three types of entities: users $u_i\in\mathcal{U}$ with $i\in\{1, \dots, N_u\}$, hashtags $h_j\in\mathcal{H}$ with $j\in\{1, \dots, N_h\}$, micro-videos $v_k\in\mathcal{V}$ with $k\in\{1, \dots, N_v\}$, and $\mathcal{U}\cup\mathcal{H}\cup\mathcal{V}=\mathcal{W}$.
Whereinto, $\mathbf{V}\in\mathbb{R}^{N_v\times D_v}$ is a feature matrix with $\mathbf{v}_k\in\mathbb{R}^{D_v}$ representing the feature vector of micro-video node $v_k$. $D_v$ is the length of the feature vector.  For the ease of presentation, $i$, $j$, and $k$ will be assigned to index user, hashtag, and micro-video, respectively. When an interaction exists between two nodes $w_i$ and $w_k$ (e.g, a user $u_i$ uploads a micro-video $v_k$), there will be an edge $e_{ik}=(w_i, w_k)\in\mathcal{E}$ to link the two nodes in the graph.
In our task, we only consider the interactions among different types of nodes, namely, user-hashtag, hashtag-microvideo, and microvideo-user. Let $\mathcal{H}_i$ denote the hashtag neighbors of user $u_i$ and $\mathcal{V}_i$ denote the micro-video neighbors of user $u_i$, namely, $\mathcal{H}_i$ is the hashtags set that $u_i$ used to tag her uploaded micro-video set $\mathcal{V}_i$.


\subsubsection{Model overview}
When users provide hashtags to describe their uploaded micro-videos, they will select the hashtags \emph{which (in their minds) can be used to describe the micro-video content that attracts them}. It implies that for the same micro-video, different users may be interested in different content and even two users are interested in the same content of a micro-video, it is still possible that they will use different hashtags due to their own preferences on hashtags. In other words, for each pair of <micro-video, hashtag>, the suitability score of the hashtag to the micro-video depends on the interests of the target user in the micro-video and personal opinions on the hashtag. In light of this, we propose a Graph Convolution Network based Personalized Hashtag Recommendation model \textbf{(GCN-PHR} for short). Given a micro-video $v_k$ uploaded by the user $u_i$, for each candidate hashtag $h_j\in\mathcal{H}$, our model will (1) generate the representations $\bar{\mathbf{v}}_k^i$ and $\bar{\mathbf{h}}_j^i$ for micro-video $v_k$ and the hashtag $h_j$ respectively based on this user $u_i$'s preferences, and then (2) compute the suitability score (or similarity score) of $h_j$ with respect to $v_k$ based on  $\bar{\mathbf{v}}_k^i$ and $\bar{\mathbf{h}}_j^i$. Therefore, the core of our model is to learn the user-specific micro-video representation  $\bar{\mathbf{v}}_k^i$  and user-specific hashtag representation $\bar{\mathbf{h}}_j^i$.


\subsection{GCN-based Representation Learning}
Intuitively, learning \emph{a user-specific micro-video representation} $\bar{\mathbf{v}}_k^i$ needs to model the interaction between  \emph{the user preference on micro-videos} $\mathbf{u_i^v}$ and \emph{the target micro-video} representation $\mathbf{v_k}$; similarly, learning \emph{a user-specific hashtag representation} $\bar{\mathbf{h}}_j^i$ has to model the interaction between \emph{the user preference on hashtags} $\mathbf{u_i^h}$ and \emph{the target hashtag} representation $\mathbf{h_j}$. Thinking one step further, the hashtags that a user used to tag micro-videos actually reflect her preference on the micro-videos to some extent; and in turn, the micro-videos that a user tagged also contain her preference on the hashtags.  In other words, $\mathcal{H}_i$ and $\mathcal{V}_i$ mutually represent each other and also reflect the user preference on each other. Therefore, it is beneficial to take  $\mathbf{u_i^h}$ into the consideration of  modeling $\bar{\mathbf{v}}_k^i$; and vice versa. In light of this, the user preference $\mathbf{u}_i$ is modeled based on both $\mathbf{u_i^v}$ and $\mathbf{u_i^h}$, and then used to learn $\bar{\mathbf{v}}_k^i$ and $\bar{\mathbf{h}}_j^i$.

Specifically, on the constructed graph the model, we adopt the message-passing idea~\cite{berg2017graph} to learn the user preference on hashtags $\mathbf{u_i^h}$ and micro-videos $\mathbf{u_i^v}$
based on the user $u_i$'s hashtag neighbors $\mathcal{H}_i$ and micro-video neighbors $\mathcal{V}_i$, respectively. After that, $\mathbf{u_i^h}$ and $\mathbf{u_i^v}$ are aggregated to learn the user preference $\mathbf{u_i}$. Finally, the $\bar{\mathbf{v}}_k^i$  is  learned based on $\mathbf{u_i}$ and $\mathbf{v}_k$, and $\bar{\mathbf{h}}_j^i$  is  learned based on $\mathbf{u_i}$ and $\mathbf{h}_j$. In the following, we introduce the learning of those representations sequentially.
\subsubsection{User preference on hashtags}
In our model, a user $u_i$'s preference on hashtags $\mathbf{u_i^h}$ is modeled by accumulating the incoming messages from all the neighbor hashtags $\mathcal{H}_i$. According to the idea of message-passing, the message transferred from a hashtag $h_j \in \mathcal{H}_i$  to the user $u_i$ is defined as:
\begin{equation}
	\mathbf{m}_{h_j \to u_i}=\mathbf{W}^u_h\mathbf{h}_j,
\end{equation}
where $\mathbf{m}_{h_j \to u_i}$ denote the message vector from hashtag $h_j$ to user $u_i$, and $\mathbf{W}^u_h$ is the weight matrix which maps the hashtag vector into the user embedding space. Based on this,  $\mathbf{u_i^h}$ is defined as:
\begin{equation} \label{eq:poh}
        \mathbf{u}_i^h = \phi(\frac{1}{|\mathcal{H}_i|}\,\sum_{h_j\in\mathcal{H}_i}\mathbf{m}_{h_j \to u_i})\,,
\end{equation}
 $\phi(\cdot)$ is the activation function and $|\mathcal{H}_i|$ denotes the number of neighbor hashtags.

\subsubsection{User preference on micro-videos}
The user preference on micro-videos $\mathbf{u_i^v}$ can be learned in the same way, namely, by accumulating the messages form all the neighbor videos $\mathcal{V}_i$. The passed message $\mathbf{v}_k$ from a micro-video represents all the contents in the video $v_k$. Notice that a micro-video contains a sequence of video frames with rich information. For a specific micro-video, a user may be only interested in its certain parts. To accurately model the user preference on micro-videos, it is crucial to identify which part in each micro-video attracts the attention of the user. Fortunately, the hashtags of a micro-video are usually provided based on the content in which that the user is interested in this micro-video. Therefore, the hashtags can better characterize the user preference on micro-videos. In light of this, in our model, the message from each video $v_k$ to a user $u_i$ is determined by its hashtags provided by the user.
Given a hashtag $h_j$ of a video $v_k$, we estimate its similarity by,
\begin{equation}
        s_{jk}=g(\mathbf{h}_j, \mathbf{W}_v^h\mathbf{v}_k),
\end{equation}
where $\mathbf{W}_v^h$ is a weight matrix and $g(\cdot)$ is a similarity function to measure the similarity of vectors. Different functions can be applied here (e.g., cosine function) and we use a fully connected layer to implement the function.

A hashtag can be used for many micro-videos by the user. Let $\mathcal{V}_{i,j}$ be the set of micro-videos tagged with the hashtag $h_j$ by user $u_i$. we normalize the similarity score $s_{jk}$ to obtain the relative similarity between a video and a hashtag based on the user preference.
\begin{equation}
        \alpha_{jk}=\frac{exp(s_{jk})}{\sum_{v_{k'}\in\mathcal{V}_{i,j}}exp(s_{jk'})},
\end{equation}
$\alpha_{jk}$ is the normalized similarity score. Let $\mathcal{H}_{i,k}$ be the hashtag set of user $u_i$ provided to micro-video $v_k$, the message of a video $v_k$ passes to a user $u_i$ is defined as:
\begin{equation}
	\mathbf{m}_{v_k \to u_i}=\sum_{h_j \in \mathcal{H}_{i,k}}\alpha_{jk}\cdot \mathbf{W}^u_v\mathbf{v}_k,
\end{equation}
It indicates that the message from $v_k$ passes to the user $u_i$ is dependent on $v_k$'s aggregated similarity to all the hashtags that  $u_i$ provides.

Similar to Eq.~\ref{eq:poh}, the user preference on micro-videos is the aggregation of the messages from all the neighbor micro-videos, namely,

\begin{equation}
        \mathbf{u}_i^v = \phi(\sum_{v_k\in\mathcal{V}_i}\mathbf{m}_{v_k \to u_i})\,.
\end{equation}
\subsubsection{User representation learning}
As discussed above, the user preference is obtained by combining the user preference on hashtags and on micro-videos. Many combination methods can be applied here, such as concatenation, addition, or more complicate deep fusion models~\cite{wang2017first}. In this work, we try two methods: network-based fusion and transformation-based summation.

\textbf{Neural network-based fusion.} In this method, $\mathbf{u}^v_i$ and $\mathbf{u}^h_i$ are first concatenated and then fed into a fully connected layer to obtain the final representation of the user preference.
Formally, the user preference is obtained by,
\begin{equation}
  \mathbf{u}_i = \phi(\mathbf{W}_{nn}\;[\mathbf{u}^v_i\ ,\ \mathbf{u}^h_i] + \mathbf{b}_{nn}),
\end{equation}
where $[\cdot, \cdot]$ is the concatenation operator;  $\mathbf{W}_{nn}$ and $\mathbf{b}_{nn}$ indicate the learnable weight matrix and bias vector in the fully connected layer, respectively. \par

\textbf{Transformation-based summation}.
In this method, $\mathbf{u}^v_i$ and $\mathbf{u}^h_i$ are fist transformed into the same space for element-wise summation.
Formally, the user preference is obtained by,
\begin{equation}
        \mathbf{u}_i = \mathbf{W}^v_u\ \mathbf{u}^v_i + \mathbf{W}^h_u\ \mathbf{u}^h_i
\end{equation}
where $\mathbf{W}^v_u$ and $\mathbf{W}^h_u$ denote the transformation matrices. \par
\subsubsection{Hashtag representation learning}
The hashtag representation is learned analogously with user preference learning. Specifically, for a hashtag $h_j$, its representation is based on both the messages from all its neighbor users and the messages from all its neighbor micro-videos. The message passed from a user to a hashtag is computed in the same way as the message passed from a hashtag to a user, and the message passed from a micro-video to a hashtag is computed in a similar way as the message passed from a micro-video to a user. To avoid the duplication of presentation, here we skip the detailed steps of how to compute the hashtag representation in our model.\par
Further, according to recent work~\cite{NGCF}, we can easily utilize message from multiple-hop neighbors by recursively stacking multiple layers to learn the representations.
\subsubsection{User-specific micro-video representation}
 The micro-video representation $\mathbf{v}_k$ is directly extracted from the content of the micro-video $v_i$, which is a concatenation of its visual, acoustic, and textual features. Detailed information about how to extract those multimodal features is described in Section 4.1.  Based on the user preference $\mathbf{u}_i$ and micro-video representation $\mathbf{v}_k$, the user-specific micro-video representation $\bar{\mathbf{v}}_k^i$ is obtained by:
\begin{equation}
        \bar{\mathbf{v}}_k^i = \phi(\mathbf{W}^v\;\mathbf{v}_k\ + \mathbf{W}^v_u\;\mathbf{u}_i+ \mathbf{b}^v),
\end{equation}
where $\mathbf{W}^v$, $\mathbf{W}^v_u$ and $\mathbf{b}^v$ denote the weight matrices and bias vector in the fully connected layer.

\subsubsection{User-specific hashtag representation}
Analogously, the user-specific hashtag representation $\bar{\mathbf{h}}_j^i$  is also learned via a fully connected layer based on the user preference $\mathbf{u}_i$ and hashtag representation $\mathbf{h}_j$,
\begin{equation}
        \bar{\mathbf{h}}_j^i  = \phi(\mathbf{W}^h\;\mathbf{h}_j\ + \mathbf{W}^h_u\;\mathbf{u}_i+ \mathbf{b}^h),
\end{equation}
where $\mathbf{W}^v$, $\mathbf{W}^v_u$ and $\mathbf{b}^v$ are the weight matrices and bias vector in the fully connected layer.

\subsubsection{Personalized Hashtag Recommendation}
Given a new micro-video $v_k$ uploaded by a user $u_i$, the hashtags in $\mathcal{H}$ could be recommended in the descending order of their similarity score with respect to $v_k$ based on based on $u_i$'s preference. Specifically, the similarity score is computed by the dot product of  the user-specific hashtag representation $\bar{\mathbf{h}}_j^i$ and the user-specific  micro-video representation $\bar{\mathbf{v}}_k^i$, namely, $(\bar{\mathbf{h}}_j^i)^T\bar{\mathbf{v}}_k^i$.
\subsection{Pairwise-based Learning}
Similar to the ranking-oriented recommendation algorithm~\cite{cao2019unifying, he2018adversarial}, we adopt the pairwise-based learning method for optimization.
To perform the pairwise ranking, it need to constructs a triplet of one micro-video $v_k$, one positive hashtag $h_j$, and one negative hashtag $h_j'$, where $h_j$ is a hashtag of $v_k$ and $h_j'$ is not. Let $\mathcal{R}= \{v_i, h_j,h_j'\}$ be the triplet sets for training. The objective function can be formulated as
\begin{equation}
	\mathop{\arg\min}_{\theta}\sum_{(\mathbf{v}_k, \mathbf{h}_j,\mathbf{h}_j')\in{\mathcal{R}}} -\ln\,\phi((\bar{\mathbf{h}}_j^i)^T\bar{\mathbf{v}}_k^i-\bar{(\mathbf{h}}_j'^i)^T\bar{\mathbf{v}}_k^i)\,+\,\lambda\left\|\Theta\right\|^2_2,
\end{equation}
where $\lambda$ and $\Theta$ represent the regularization weight and the parameters of the model, respectively.
\section{Experiments}
In this section, we first present the experimental settings~(i.e. datasets, baselines, evaluation protocols, and parameter settings), and then report and analyze the experimental results. Finally, we give some visualization examples.
\vspace{-1mm}
\begin{table}
  \centering
  \caption{Statistics of the evaluation dataset.~(\#Micro-videos, \#Hashtags, and \#Users denote the numbers of micro-videos, Hashtags, and users, respectively.)}
    \vspace{-3mm}
  \label{table_1}
  \setlength{\tabcolsep}{2.0mm}
  \begin{tabular}{c|c|c|c}
    \hline
    Dataset&\#Micro-videos&\#Users&\#Hashtags\\
    \hline
    YFCC100M&134,992&8,126&23,054\\
    \hline
    Instagram&48,888&2,303&12,194\\
    \hline
  \end{tabular}
  \vspace{-2mm}
\end{table}

\begin{table}
  \centering
  \caption{Feature summarization of three modalities.~(Visual, Acoustic, and Textual denote the dimensions of visual, acoustic, and textual modalities, respectively.)}
    \vspace{-3mm}
  \label{table_2}
  \setlength{\tabcolsep}{4mm}
  \begin{tabular}{c|c|c|c}
    \hline
    Dataset&Visual&Acoustic&Textual\\
    \hline
    YFCC100M&2,048&128&100\\
    \hline
    Instagram&2,048&128&100\\
    \hline
  \end{tabular}
  \vspace{-3mm}
\end{table}

\begin{table*}
  \centering
  \setlength{\tabcolsep}{1.5mm}
  \caption{Performance comparison between our model and the baselines.}
    \vspace{-3mm}
  \label{table_3}
  \begin{tabular}{c|cc|cc|cc|cc|cc|cc}
    \hline
    \multirow{2}{*}{Model}&\multicolumn{6}{c|}{YFCC100M}&\multicolumn{6}{c}{Instagram}\\
    \cline{2-13}&P@5&P@10&R@5&R@10&A@5&A@10&P@5&P@10&R@5&R@10&A@5&A@10\\
    \hline
    UTM&0.2469&0.1780&0.3053&0.4108&0.6860&0.7724&0.5193&0.4503&0.2305&0.3659&0.7823&0.8136\\
    ConTagNet&0.3595&0.2441&0.4358&0.5426&0.7856&0.8395&0.4391&0.3863&0.1982&0.3136&0.74.44&0.8178\\
    CSMN&0.3027&0.1897&0.4004&0.4428&0.6403&0.6634&0.5230&0.4691&0.2614&0.3826&0.8246&0.8770\\
    USHM&0.3891&0.2711&0.4865&0.6158&0.8398&0.8877&0.6613&0.5821&0.3402&0.5134&0.9107&\textbf{0.9365}\\
    GCN-PHR&\textbf{0.4004}&\textbf{0.3053}&\textbf{0.5125}&\textbf{0.6667}&\textbf{0.8403}&\textbf{0.9021}&\textbf{0.6847}&\textbf{0.6286}&\textbf{0.4075}&\textbf{0.5667}&\textbf{0.9134}&0.9332\\
    \hline
    \hline
    Improv.&\textbf{2.90\%}&\textbf{12.61\%}&\textbf{5.34\%}&\textbf{8.30\%}&\textbf{0.01\%}&\textbf{1.62\%}&\textbf{3.54\%}&\textbf{7.99\%}&\textbf{19.78\%}&\textbf{10.38\%}&\textbf{0.30\%}&-0.35\%\\
    \hline
  \end{tabular}
  \vspace{-2mm}
\end{table*}
\begin{table*}
  \centering
  \setlength{\tabcolsep}{1.5mm}
  \caption{Performance comparison between our model and the variants.}
    \vspace{-3mm}
  \label{table_4}
  \begin{tabular}{c|cc|cc|cc|cc|cc|cc}
    \hline
    \multirow{2}{*}{Model}&\multicolumn{6}{c|}{YFCC100M}&\multicolumn{6}{c}{Instagram}\\
    \cline{2-13}&P@5&P@10&R@5&R@10&A@5&A@10&P@5&P@10&R@5&R@10&A@5&A@10\\
    \hline
    \textbf{Variant-\uppercase\expandafter{\romannumeral1}}&0.3786&0.2745&0.4923&0.6266&0.8335&0.8901&0.6703&0.58330.&3473&0.5232&0.9068&0.9245\\
    \textbf{Variant-\uppercase\expandafter{\romannumeral2}}&0.3829&0.2837&0.5031&0.6401&0.8307&0.8834&0.6765&0.6053&0.3735&0.5398&0.9089&0.9306\\
    \textbf{Variant-\uppercase\expandafter{\romannumeral3}}&0.3801&0.2754&0.4966&0.6312&0.8283&0.8831&0.6725&0.5974&0.3665&0.5274&0.8953&0.9268\\
    GCN-PHR&\textbf{0.4004}&\textbf{0.3053}&\textbf{0.5125}&\textbf{0.6667}&\textbf{0.8403}&\textbf{0.9021}&\textbf{0.6847}&\textbf{0.6286}&\textbf{0.4075}&\textbf{0.5667}&\textbf{0.9134}&\textbf{0.9332}\\
    \hline
  \end{tabular}
  \vspace{-3mm}
\end{table*}

\subsection{Experimental Settings}
\textbf{Datasets.}
We conducted experiments on a public dataset YFCC100M \cite{veit2018separating} and a self-collected Instagram dataset.
The characteristics of these datasets are summarized in Table~\ref{table_1}.

\textbf{YFCC100M.}\footnote{https://multimediacommons.wordpress.com/yfcc100m-core-dataset/.}
  The YFCC100M dataset is the largest publicly accessible multimedia collection, containing the metadata of around 99.2 million photos and 0.8 million videos from Flickr.
  In our task, we focus on the personalized micro-video hashtag recommendation. Therefore, we crawled video dataset, user profiles, and annotated hashtags based on its API.
  Finally, a dataset with $134,992$ micro-videos, $8,126$ users, and $23,054$ hashtags are collected.
  After that, we extracted a rich set of features from textual, visual and acoustic modalities.
  Specially, we adopted FFmpeg\footnote{https://ffmpeg.org/.} to extract the keyframes from micro-videos, and then employed the ResNet50~\cite{he2016deep} model pre-trained by Pytorch\footnote{https://pytorch.org/.} to extract the visual features.
  Simultaneously, we separated audio tracks from micro-videos with FFmpeg, and adopted VGGish~\cite{hershey2017cnn} to learn the acoustic deep learning features.
  In addition, we utilized Sentence2Vector~\cite{arora2016simple} trained with twitter text dataset to extract the textual features from micro-video descriptions.

 \textbf{Instagram.}
  To construct a micro-video dataset for evaluating our method, we crawled the micro-video associated with the description from Instagram.
  We randomly started with some users and iteratively crawled the list of users who followed those users.
  In such a way, we harvested about 1 million micro-videos, 10 thousand users, and 100 thousand hashtags.
  To evaluate our method, we removed the micro-videos without any hashtag, and cleaned the users whose  micro-videos are all discarded accordingly.
  After the pre-processing, we obtained a micro-video dataset of $48,888$ micro-videos, $2,303$ users, and $12,194$ hashtags.
  We perform a similar process to extract the features of micro-videos in the Instagram dataset. 
 For further clarification, we summarized the multimodal features of two datasets in Table~\ref{table_2}.

\noindent\textbf{Baselines}.
To evaluate the effectiveness of our model, we compared our proposed method with several state-of-the-art baselines.\par
\begin{itemize}
  \item \textbf{UTM}~\cite{denton2015user}.
  The method utilizes a 3-way gating to combine heterogeneous features into a learning framework where the model is conditioned on the user profile. This model learns a joint $d$-dimensional embedding space for posts and hashtags.
  \item \textbf{ConTagNet}~\cite{rawat2016contagnet}.
 This baseline proposes a convolutional neural network~(CNN) based framework to integrate user profiles with posts' content for hashtag prediction.  In the experiment, for the fairness, we leveraged the extracted multimodel information and fed them into the fully connected network with the corresponding users' information to estimate the hashtags.
  \item \textbf{CSMN}~\cite{chunseong2017attend}. The method named as Context Sequence Memory Network~(CSMN) which leverages the historical information as the prior knowledge of users' vocabularies or writing styles. Towards this end, the content information extracted from micro-videos and hashtags users annotated is used as the memory retaining the context information. Besides, previously generated hashtags can be appended into memory to capture long-term information. With the proposed CNN memory structure, hashtags which meet the context information can be predicted.
  \item \textbf{USHM}~\cite{veit2018separating}.
    This approach develops a user-specific hashtag model that takes the hashtag usage patterns of a user into account, and adopts a three-way tensor model to learn user embeddings. This baseline trains the model on image-user-hashtag triplets that allow the model to learn patterns in the hashtag usage of particular users and to disambiguate the learning signal.
\end{itemize}
\textbf{Evaluation Protocols and Parameter Settings.}
We randomly split the dataset into training, validation, and testing sets with 8:1:1 ratio as in~\cite{cheng2016effective2,cheng2018aspect,cheng2019mmalfm}, and created the training triples based on the random negative sampling strategy.
The commonly used metrics \textit{precision}, \textit{recall}, and \textit{accuracy} are used for evaluation.
To train our proposed model, we use the $leaky\_relu$ as the activation function in our model and randomly initialize model parameters with a Gaussian distribution. The stochastic gradient descent~(SGD) algorithm is used for  optimization.
We have tried different parameter setings, including the batch size of $\{128, 256, 512\}$, the latent feature dimension of $\{32, 64, 128\}$, the learning rate of $\{0.0001, 0.0005, 0.001. 0.005, \\ 0.01\}$ and the regularizer of $\{0, 0.00001, 0.0001, 0.001, 0.01, 0.1\}$.
In addition, we sample $1,000$ negative hashtags for each micro-video in the test process.
As the findings are consistent across the dimensions of latent vectors, if not specified, we only report the results based $D=64$, which gives relatively  good performance.
\vspace{-1mm}
\subsection{Performances Comparison}
We report the results of different methods in Table~\ref{table_3}.
From the results, we have the following observations.
Firstly, our proposed method achieves the best performance across  all the three datasets, demonstrating the effectiveness of our model. This also verifies the importance of modeling  user preference and hashtag semantic  by considering the micro-video content information.

Secondly, USHM outperforms all the other baselines besides our model. This  method captures the hashtag usage pattern of the users based on the historical interactions between user and hashtag, indicating the necessity of considering user preference in hashtag recommendation.
However, because USHM merely models users' interactions between hashtags and posts while ignores post contents, it is still inferior to our model, which models the interaction among users, hashtags, and posts.

In addition, we notice that our method achieves a bigger improvement on YFCC than that on Instagram in terms of precision, while it is just the opposite in terms of recall. This is mainly because the average number of  hashtags  per  micro-video in Instagram is larger than that of in YFCC.
In terms of accuracy, our method does not have significant improvement over USHM or even lower in some cases.
The reason is that the accuracy calculates the ratio of at least one ground truth hashtag appears in the top $K$ results\footnote{{We adopt the definition of accuracy in~\cite{veit2018separating}, which is used to measure the ratio of the micro-videos annotated correctly by at least one hashtag.}}.

Finally, it is expected that the performance of CSMN is the worst on YFCC. Remind that CSMN represents users based on their previous used hashtags for hashtag recommendation. Therefore, it is more sensitive to the density of the dataset. For datasets of high sparsity, its performance becomes unsatisfactory.
\vspace{-2mm}
\subsection{Ablation Study}
In this section, we studied the effectiveness of each component of our proposed model by comparing the following variants.

\begin{itemize}
\item \textbf{Variant-\uppercase\expandafter{\romannumeral1.}} In this model, we built a user-hashtag-microvideo graph and performed graph convolution operations on each node without the attention mechanism. This variant is designed to investigate the effectiveness of attention mechanism in our model.
\item \textbf{Variant-\uppercase\expandafter{\romannumeral2.}} To evaluate the effectiveness of the user preference to hashtag representation, we discard the user preference on micro-video content and passed the  information from micro-video nodes to the hashtag nodes.
\item \textbf{Variant-\uppercase\expandafter{\romannumeral3.}} In contrast to \textbf{Variant-\uppercase\expandafter{\romannumeral2.}}, this variant removes the guidance of hashtags on the micro-video feature extraction and it aggregates the micro-video content information directly. This method is to evaluate the importance of hashtags on user preference modeling.
\end{itemize}
\vspace{-1mm}
The results of those variants and our method are shown in Table~\ref{table_4}. As expected, GCN-PHR outperforms all the variants by a large margin, especially on the Instagram dataset. This verifies our assumption on personalized hashtag recommendation: we should consider both user preferences on the content of posts (i.e., micro-video here) and user preferences on hashtags. They are mutually influence each other on the use of hashtags. It also demonstrates the effectiveness of our strategy on learning the user and hashtag representations.  From the performance of the three variants, we can see that the user preference and hashtag representations are improved by content information. In addition, the user preference and hashtag semantic can help the feature extraction during the message-passing, facilitating the hashtag and user preference representation learning. \par

 The performance of \textbf{Variant-\uppercase\expandafter{\romannumeral1}} is the worst, which demonstrates the importance of using attention mechanism to identify user's interested contents in micro-videos and user preference on hashtags. It also implies the effectiveness of our strategy by using the hashtags (or micro-videos) to attend the video (or hashtag) content for representation learning.   From the performance of \textbf{Variant-\uppercase\expandafter{\romannumeral2}} and
    \textbf{Variant-\uppercase\expandafter{\romannumeral3}}, we can find that (1) it is very important to consider user preference (on micro-videos) to learn the hashtag representations for personalized recommendation; and (2) it is even more important to use hashtag to learn user preference representation (as the performance of \textbf{Variant-\uppercase\expandafter{\romannumeral2}} is better than \textbf{Variant-\uppercase\expandafter{\romannumeral3}}. Notice that the hashtag is used to help identify users' interested content from the abundant information of micro-videos, which is crucial to learn user preference.


\begin{figure}
	  \includegraphics[width=0.48\textwidth]{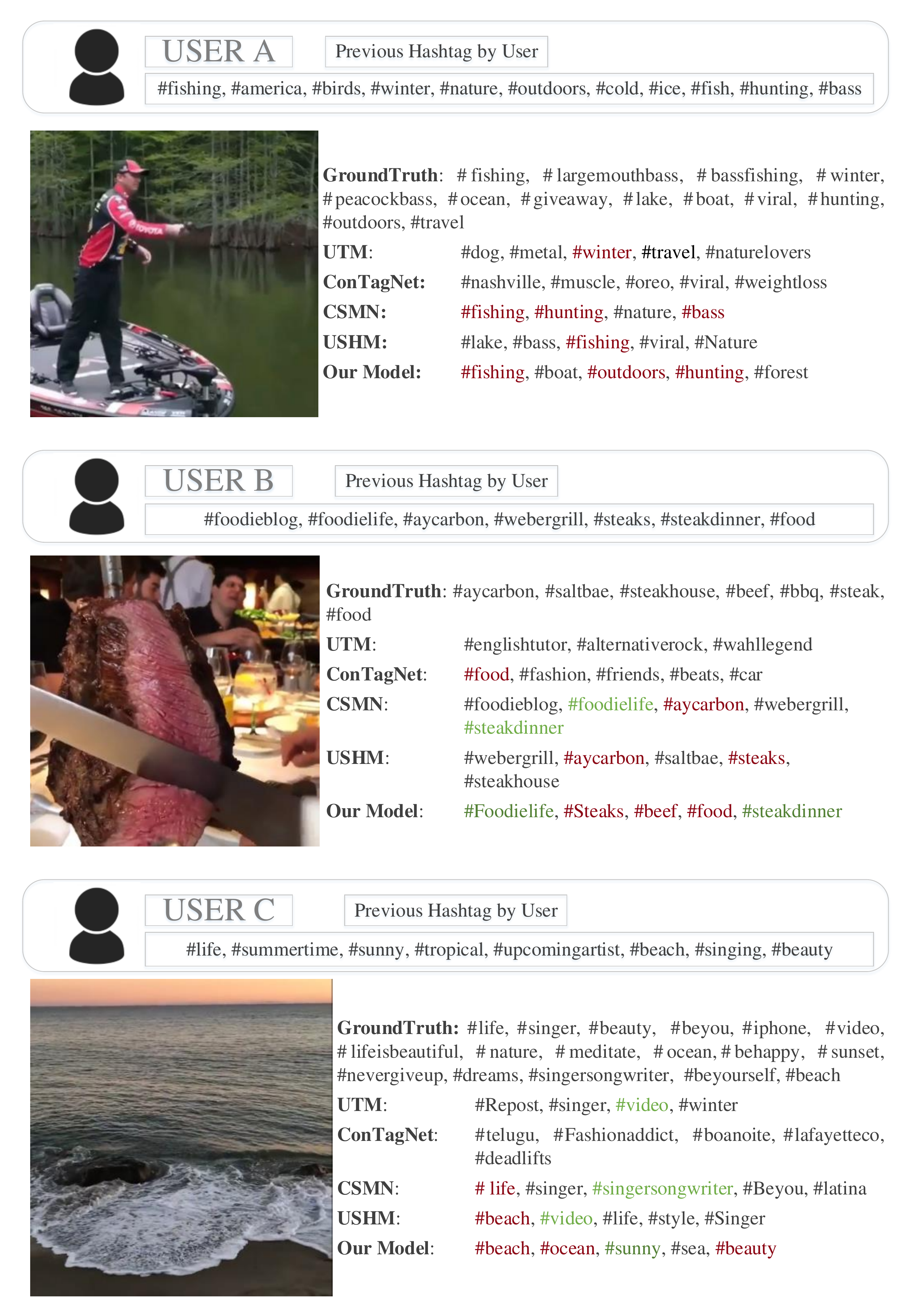}
	  \vspace{-4mm}
	  \caption{Visualization of hashtag recommended by baselines and our model on the Instagram dataset. }
	\label{fig_3}
	\vspace{-4mm}
\end{figure}
\subsection{Visualization}
In this section, we randomly selected a number of micro-videos to visually validate the effects of our model, taking three specific micro-videos as examples. For each micro-video, in this figure, we show \textit{the hashtags that the corresponding user provided}, \textit{the recommended hashtags by different methods}, and \textit{the historical hashtag used by the user}. The hashtags in red are the ones predicted correctly, the ones in green are used in the past but have not been used in this micro-video.

In the first example, it can be seen that our method can accurately recommend the hashtags of `\#fishing', `\#hunting' `\#boat', and `\#outdoors'. In these hashtags, `\#hunting', `\#fishing', and `\#outdoors', which are all the hashtags used in the past. This indicates our method can capture user's hashtag usage preference. USHM and CSMN also recommend `\#fishing', `\#hunting', and `\#outdoors' based on the hashtag usage patterns of the user. However, these methods miss the hashtag of `\#boat', which has not been used by the users before.  It is impossible to recommend such hashtags merely based on modeling user's hashtag usage habits. Different from these methods, our method takes content information into considerations to recommend suitable hashtags based on user's preference on the current micro-video's content and her preference on hashtags. Therefore, our model can successfully recommend the hashtags which have not been used by the users before.

As shown in the second example, our model predicts hashtags `\#steaks', `\#beef', and `\#food' to the micro-video accurately, in which `\#beef' has not been used by users.
    This also demonstrate the benefit of using the content information in hashtag recommendation. Notice that although the recommended `\#foodielife' and `\#steakdinner' are not in the ground-truth,  they are actually fit for the micro-video content and user's hashtag usage pattern. This also demonstrates the potential of our model on suggesting hashtags.

In the last example, our presented model recommends the third micro-video with `\#beach', `\#ocean', and `\#beauty'. Different from other hashtags, `\#beauty' does not have a specific visual appearance. And different users have their own tastes on `beauty'. However, our model can still correctly recommend this hashtag, this further validates the powerful capability of our model on learning user-specific hashtag semantic and matching it with the user-interested micro-video contents.

Overall, our proposed method outperforms all the baselines, especially for the micro-videos with many concepts. This indicates that our proposed model can effectively learn the micro-video and hashtag representation based on user preference and recommend proper personalized hashtags accordingly.

\section{conclusion}
When users provide hashtags for their uploaded micro-videos, they would like to use the ones which, in their mind, can describe the content that they are interested in. Because users' distinct preferences on video contents and their own hashtag usage habits or understanding, even for the same micro-video, they may use very different hashtags. This observation indicates that it is important to model the complex interactions among users, hashtags, and micro-videos for accurate hashtag recommendation. In this paper, we present a Graph Convolution Network based Personalized Hashtag Recommendation (GCN-PHR) model to achieve the goal. In particular, our model leverages the GCN techniques (the message-passing model and attention mechanism) to learn the representations of  micro-videos and  hashtags based on user preference. The learned user-specific micro-video representation and user-specific hashtag representation are used to compute the suitability score (or similarity score) of the hashtag with respect to the micro-video for hashtag recommendation.
To demonstrate the effectiveness of our model, extensive experiments have been conducted on two micro-video benchmark datasets. The experimental results show that our proposed model outperforms several state-of-the-art baselines.
In addition, we also conduct ablation studies to analyze the effectiveness of different components of our model and use case studies to illustrate the performance.

\section{Acknowledgments}
This work is supported by the National Natural Science Foundation of China, No.: 61772310, No.:61702300, No.:61702302, No.: 61802231, and No. U1836216; the Project of Thousand Youth Talents 2016; the Shandong Provincial Natural Science and Foundation, No.: ZR2019JQ23,  No.:ZR2019QF001; the Future Talents Research Funds of Shandong University, No.: 2018WLJH 63.
\bibliographystyle{ACM-Reference-Format}
\bibliography{HashtagRec}

\end{document}